\documentclass{article}
\usepackage[nonatbib,final]{mlncp_2023}
\usepackage[numbers]{natbib}
\usepackage{amsmath,amssymb,colortbl,array}
\usepackage{graphicx} % Required for inserting images

\graphicspath{{Figures/}}
\usepackage[subpreambles=true]{standalone}

\usepackage[para]{footmisc}
\usepackage[misc]{ifsym}

\usepackage{dcolumn}
\usepackage{xcolor}
\usepackage{fancyhdr}
\usepackage{lipsum}
\usepackage{soul}
\usepackage{braket}
\usepackage{enumitem}
\usepackage{titlesec}

\usepackage[colorlinks=true,citecolor=blue,linkcolor=blue,urlcolor=blue]{hyperref}

\usepackage{lmodern}
\usepackage{booktabs}
\usepackage{footnote}
\usepackage{lettrine}
\usepackage[ruled,vlined,linesnumbered]{algorithm2e}
\makeatletter
\newcommand*{\addFileDependency}[1]{% argument=file name and extension
  \typeout{(#1)}
  \@addtofilelist{#1}
  \IfFileExists{#1}{}{\typeout{No file #1.}}
}
\makeatother

\title{Mean-Field Assisted Deep Boltzmann Learning with Probabilistic Computers}
\author{Shuvro Chowdhury, Shaila Niazi, Kerem Y. Camsari \\ \\ Department of Electrical and Computer Engineering\\ University of California Santa Barbara, Santa Barbara, CA 93106, USA \\ \{schowdhury, sniazi, camsari\}\,@\,ucsb.edu}
\date{September 2023}
\vspace{-20pt}
\begin{document}

\maketitle
\begin{abstract}

Despite their appeal as physics-inspired, energy-based and generative nature, general Boltzmann Machines (BM) are considered intractable to train. This belief led to simplified models of BMs with \emph{restricted} intralayer connections or layer-by-layer training of deep BMs. Recent developments in domain-specific hardware -- specifically probabilistic computers (p-computer) with probabilistic bits (p-bit) -- may change established wisdom on the tractability of deep BMs. In this paper, we show that deep and \emph{unrestricted} BMs can be trained using p-computers generating hundreds of billions of Markov Chain Monte Carlo (MCMC) samples per second, on sparse networks developed originally for use in D-Wave's annealers. To maximize the efficiency of learning the p-computer, we introduce two families of Mean-Field Theory assisted learning algorithms, or xMFTs (x = Naive and Hierarchical). The xMFTs are used to estimate the averages and correlations during the \emph{positive phase} of the contrastive divergence (CD) algorithm and our custom-designed p-computer is used to estimate the averages and correlations in the negative phase. A custom Field-Programmable-Gate Array (FPGA) emulation of the p-computer architecture takes up to 45 billion flips per second, allowing the implementation of CD-$n$ where $n$ can be of the order of millions, unlike RBMs where $n$ is typically 1 or 2. Experiments on the full MNIST dataset with the combined algorithm show that the positive phase can be efficiently computed by xMFTs without much degradation when the negative phase is computed by the p-computer. Our algorithm can be used in other scalable Ising machines and its variants can be used to train BMs, previously thought to be intractable. 
\end{abstract}

\vspace{-5pt}
\section{Introduction}
\vspace{-10pt}
Since their introduction by Hinton and colleagues \cite{ackley1985learning}, Boltzmann Machines (BM) have received sustained interest over the years \cite{salakhutdinov2009deep}. Most recently, BMs have found renewed interest in the representation of quantum many-body wavefunctions as an alternative to quantum Monte Carlo algorithms (see, for example, \cite{carleo2019machine}). Meanwhile, the nearing end of Moore's Law has been driving the development of domain-specific computers tailored for specific applications and algorithms. A notable class of such computers deals with probabilistic computers (p-computer) with probabilistic bits (p-bit) (see, for example, \cite{camsari2017stochasticL,chowdhury2023full}). Probabilistic computers have been implemented at various sizes and in different physical substrates. Magnetic nanodevices exploiting the ambient thermal noise to build p-bits have been implemented in small scales (10-80 p-bits, \cite{borders2019integer,si2023energy}). Custom-made digital \emph{emulators} using Field Programmable Gate Arrays (FPGA) has been scaled much further, up to 5,000 - 10,000 p-bits \cite{Aadit2022a}. Despite their small sizes at table-top experiments, nanodevice-based p-computers are arguably the most scalable option to gain more performance and energy-efficiency, with projections up to million p-bit densities \cite{sutton2020autonomous}, given the success of magnetic memory technology \cite{lin200945nm}. The high costs of pseudorandom number generators required for each p-bit make it prohibitively expensive to get to such million-bit densities using existing CMOS technology \cite{kobayashi2023cmos+}.  
\begin{figure}[!t]
    \vspace{0pt}
    \centering
    \includegraphics[width=0.95\textwidth,keepaspectratio]{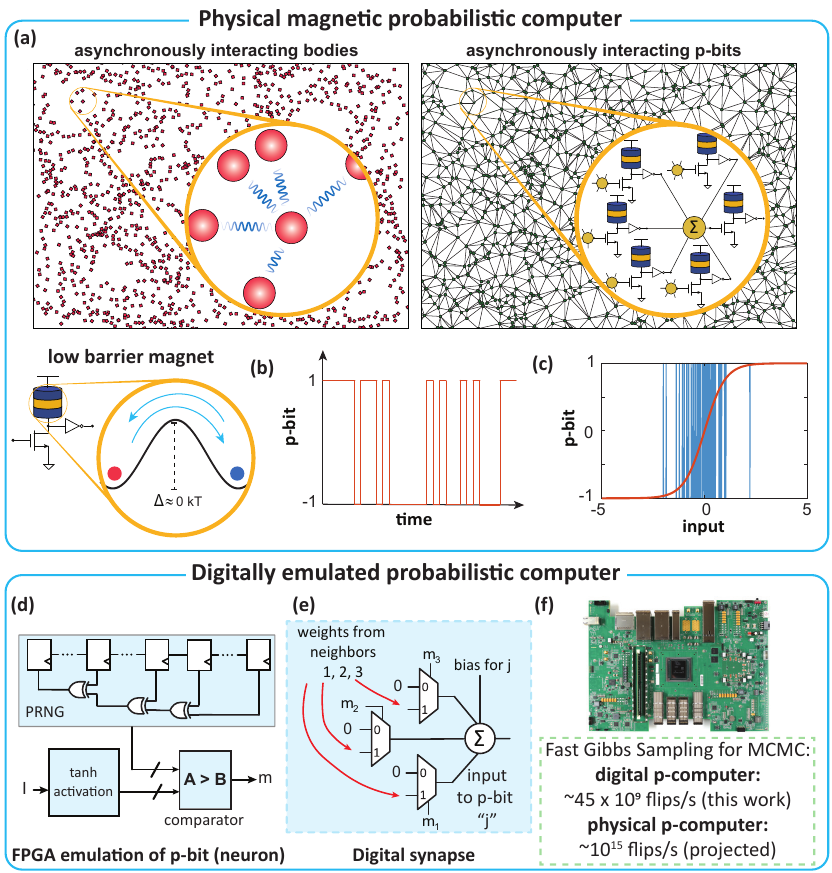}
    \caption{\textbf{p-computing overview:} (a) Analogy between interacting bodies in nature and interacting p-bit networks we build in this work. In stochastic MTJ (sMTJ) based implementations of p-bits, a low energy barrier magnet is used to generate natural noise. (b) Typical output of a p-bit against time fluctuating randomly between $+1$ and $-1$. (c) Input/output characteristic of a p-bit. The output (blue curve) is pinned to $\pm 1$ at strong positive and negative inputs. The average (orange) has a tanh behavior. (d) In this work, we emulate the p-bit in a digital system (FPGA) with a pseudorandom number generator (PRNG), a lookup table for the tanh and a comparator. (e) The digital emulation of the synapse with MUXes is also shown. (f) A p-computer consisting of a network of such p-bits is then realized in an FPGA. }
    \label{fig:fig1}
    \vspace{-20pt}
\end{figure}
Nonetheless, CMOS emulators of p-computers are useful in demonstrating the architectural and algorithmic potential of physics-inspired scalable p-computers. In this paper, we use such a custom-designed, highly efficient FPGA-based p-computer emulator (Fig.~\ref{fig:fig1}d-f) that can take 45 billion MCMC samples every second (or flips per second, fps) to train deep Boltzmann machines. Notably, this flips per second is about 4 - 5X faster than custom GPU/TPU implementations implemented on far simpler networks with $\pm 1$ weights (see, for example, \cite{adachi2015application,manukian2019accelerating,dixit2021training,bohm2022noise}). Boltzmann Machines are trained by the contrastive divergence algorithm, assuming a quadratic energy function \cite{hinton2002training,larochelle2007empirical},%\vspace{-5pt}
\begin{equation} 
    \Delta W_{ij}  = \langle m_i m_j \rangle^{\mbox{data}}-\langle m_i m_j \rangle^{\mbox{model}} \qquad \text{and} \qquad \Delta h_i = \langle m_i  \rangle^{\mbox{data}}-\langle m_i \rangle^{\mbox{model}}
    \label{eq:CD}
\end{equation}

%\vspace{-5pt}
One difficulty in training unrestricted Boltzmann machines is the need to perform explicit MCMC sampling in the positive (data) phase. To perform the visible-to-hidden layer inference in one shot, BMs typically restrict connections within a layer, removing the need for MCMC sampling. Removing these connections, however, hurts the representational ability of the network. The need for two separate phases is also detrimental to hardware development \cite{niazi2023training} where each input in a batch needs to be clamped followed by MCMC sampling, \emph{serially}. In this paper, we propose a hybrid algorithm to circumvent the positive phase sampling of \emph{unrestricted} Boltzmann machines. Our main contributions are as follows:

(1) We implement a fast FPGA-based digital MCMC sampler emulating physical probabilistic computers that are able to take up to 45 billion Gibbs samples per second, communicating with a classical computer in a closed-loop setup capable of training a deep \emph{unrestricted} BM with 2560 nodes and 17984 parameters to learn the full MNIST data set entirely in hardware, which is rarely performed in direct hardware implementations of BMs.  

(2) We propose a hybrid mean-field theory (MFT) assisted contrastive divergence (CD) algorithm to ease the positive phase computation of \emph{unrestricted} and \emph{deep} BMs. Going beyond naive MFTs (NMFT), we also propose a \emph{hierarchical} MFT (HMFT), improving correlation estimations at the cost of making $\mathcal{O}(N^2)$ more NMFT calls. \vspace{3pt}

(3) We demonstrate that the hybrid algorithm we design does not result in significant degradation compared to the MCMC method since \emph{positive} phase correlations are much more easily handled by MFTs as opposed to \emph{negative} phase correlations.  \vspace{3pt}

%FIG. 2 goes about here. 
\begin{figure}[t]
    \vspace{0pt}
    \centering    \includegraphics[width=0.99\textwidth,keepaspectratio]{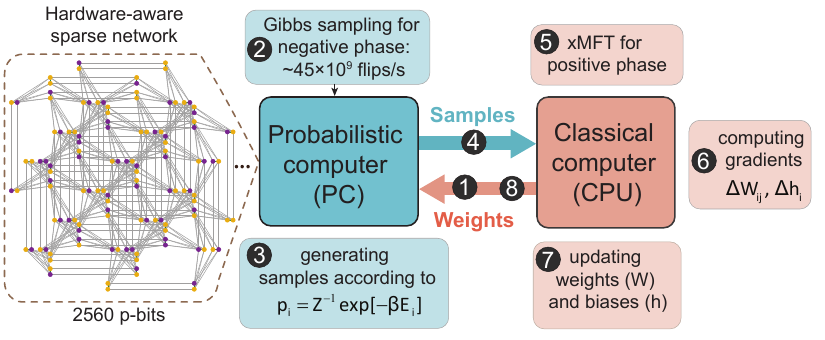}
    \caption{\textbf{Hybrid computing scheme for ML:} A hybrid computing scheme with probabilistic and classical computers is shown. Inside the classical computer, the positive phase is performed with the help of mean-field theory derivative algorithms. At the beginning of the negative phase, the classical computer sends weights and biases required to our probabilistic computer (PC) where we perform Gibbs sampling. The probabilistic computer can generate a measured 45 billion Gibbs flips in a second (FPGA). The PC returns samples to the CPU which computes the gradient. This process is repeated until convergence.}
    \label{fig:fig2}
    \vspace{-10pt}
\end{figure}

%\subsection{p-computing fundamentals}
\vspace{-5pt}
\section{Gibbs Sampling with p-bits and Mean Field Theories}
\vspace{-10pt}
A p-bit randomly fluctuates between two states (say, in between $+1$ and $-1$) with a continuous-valued input. Mathematically, an interconnected network of p-bits is represented by the following two equations:
\begin{align}
I_i = \sum_{j}{W_{ij}m_j}+h_i \qquad \text{and}\qquad m_i=\mathrm{sgn}\left(\tanh{\left(\beta I_i\right)-r_{[-1,1]}}\right) \label{eq:pbit}
\end{align}\vspace{-15pt}

where $m_i\in \{-1,+1\}$ and $r_{[-1,1]}$ is a uniform random number drawn from the interval $[-1,1]$. $\{W_{ij}\}$ are the weights, $\{h_i\}$ are the biases and $\beta$ is the inverse temperature. When solved iteratively also known as Gibbs sampling \cite{Koller_probModels}, Eq.~(\ref{eq:pbit}) approximately follows the Boltzmann distribution \cite{camsari2017stochastic}:
\begin{align}
p(\{m\}) = \frac{1}{Z}\,\exp{[-\beta E(\{m\})]} \qquad\text{and}\qquad E(\{m\})=-\sum_{i<j}W_{ij}m_im_j-\sum_{i}h_im_i
\end{align}
The second equation in Eq.~(\ref{eq:pbit}) is also known as the ``binary stochastic neuron'' in machine learning. In the present context, the iterated evolution of these equations represents a \emph{dynamical system} directly implemented in hardware. As long as $I_i$ computation time is faster than $m_i$  computation time in Eq.~(\ref{eq:pbit}), the update order does not matter and can be random, supporting asynchronous and massively parallel designs. In general-purpose computers where Gibbs sampling is usually performed on software, it can become computationally expensive, especially without the help of any accelerator. Therefore, in many fields of physics and statistics, mean-field theory (MFT) is instead widely used where one tries to approximate the behavior of a many-body system by using an average field instead of individual interactions among the components \cite{chaikin_lubensky_1995},\cite{kardar_2007}. This simplification of the system description to a mere average significantly reduces the computational load.  In the present context, the relevant MFT equations to be solved self-consistently are the following \cite{Dalton_magnetisationMFT} where $\langle m \rangle \in (-1,1)$:
\begin{align}
\langle I_i \rangle = \sum_{j}{W_{ij} \langle m_i \rangle}+h_i  \qquad \text{and} \qquad \langle m_i \rangle = \tanh{ (\beta\langle I_i \rangle) }\label{eq:MFT}
\end{align}
It is also worthwhile to note that although MFT yields a solution of a complex system with less computational effort, the estimates from MFT are not always accurate \cite{Dalton_magnetisationMFT}. 
\vspace{-5pt}
\section{Hierarchical Mean Field Assisted CD Algorithm}
\vspace{-10pt}
In the context of Boltzmann machine learning, the idea of replacing the correlations in Eq.~(\ref{eq:CD}) with MFTs was first introduced by Petersen et al. \cite{petersen_1987_MFT_learning}. Improvements to this idea such as linear response correction \cite{Kappen_advances_1997,KAPPEN1997_MFTapproach,kappen1998efficient} and its higher order extensions \cite{Tanaka1998MFT} were also made. Hinton et al. \cite{Welling2002_newLearning} proposed a deterministic variant of the contrastive divergence algorithm. Recently, a variational mean field theory has also been proposed in \cite{Hang2020variational}. 

Different from all these approaches considered earlier, in this paper we propose a hybrid approach (Fig.~\ref{fig:fig2}): unlike most MFT approaches, the free running phase is performed with Gibbs sampling but on a fast p-computer and for the positive phase in the spirit of \cite{salakhutdinov2009deep}, we use an alternative method to compute the correlations from the vanilla MFT method which we call hierarchical mean-field theory (HMFT). In traditional MFT methods, the correlations are calculated assuming independence between interacting bodies, i.e., $\langle m_im_j\rangle=\langle m_i\rangle \langle m_j\rangle$ \cite{kappen1998efficient}. In our approach, we do not use this assumption. Rather, we start from the basic definition of correlation, i.e.,
\begin{align}
\langle m_im_j\rangle &=\sum_{m_i=\pm 1, m_j = \pm 1} p(m_i,m_j) m_i m_j \qquad \text{with} \qquad p(m_i,m_j) =  p(m_i|m_j)\,p(m_j)
\label{eq:correlation_basic}
\end{align}
When we compute MFT, we get an estimate for $p(m_j)$. However, to use Eq.~(\ref{eq:correlation_basic}), we also need to know or compute $p(m_i|m_j)$. This can be done by \emph{clamping} p-bit $j$ to $\pm 1$ and then performing another MFT estimating $p(m_i|m_j)$. After making $\Theta(2n)$ such MFT calls, we can estimate the second-order correlations using Eq.~\eqref{eq:correlation_basic}. Our HMFT approach is presented in Algorithm~\ref{alg:alg3}. HFMT improves correlation estimations by not baking in the independence assumption $\langle m_i m_j \rangle = \langle m_i \rangle \langle m_j \rangle$ as Naive MFT does (see Supplementary Information~\ref{sec:toy_example} for a discussion on how HMFT can capture correlations completely missed by MFT assuming independence, in a toy example).  In fact, this Bayesian trick can be used in conjunction with other methods that approximate marginals, such as Loopy Belief Propagation \cite{mezard2009information} or with corrections to naive MFT (for example, see \cite{Kappen_advances_1997}), improving the HFMT method. Moreover, higher-order correlations of the form $\langle m_i m_j m_k \rangle$, $\langle m_i m_j m_k m_l \rangle$, $\ldots$ can be \emph{hierarchically} estimated. These can then be used to train higher-order Boltzmann machines \cite{sejnowski1986higher},  trading off parallelizable MFT computations with the fundamentally serial Gibbs sampling. In the experiments below, we investigate how the positive phase of the contrastive divergence algorithm could be performed by the MFT and the HFMT method we propose. 
\vspace{-10pt}
% please don't delete the following lines
\SetKwComment{Comment}{\normalfont $\triangleright$ }{}
\SetKwInOut{Input}{Input}
\SetKwInOut{Output}{Output}
\SetKwInput{KwSampler}{Sampler}
\SetKw{KwBy}{By}
\begin{algorithm}[!ht]
\DontPrintSemicolon
\caption{\textbf{The Hierarchical Mean-field Algorithm}}
\label{alg:alg3}
\Input{weights and biases $J, h$, update factor $\lambda$, tolerance $\delta$, max. iteration $T_{\mathrm{max}}$}
%\KwSampler{Classic Gibbs CPU, Graph-colored Gibbs CPU, Graph-colored Gibbs FPGA}
\Output{estimates for averages $\langle m_{\text{i}}\rangle$ and correlations $\langle m_{i}m_{j} \rangle$}
$\epsilon \gets 1000$,\quad $N \gets  \text{length}(h)$,\quad $T \gets  1$,\quad $m_{\text{old}} \gets 0.01\,\text{rand}(-1,1)$\;
\For{$i\gets 1$ \KwTo $N+1$}{
    \For{$j \gets -1 $ \KwTo $+1$ \KwBy $2$}{
        \If{$i\neq 1$}{  
            $m_{\text{old},i-1} \gets j$ \Comment*[r]{clamping to $\pm 1$ to get conditional probability}
        }
        %\tcc{mean field iterations}
        \While{$\epsilon \geq \delta$}{
                $I \gets Jm_{\text{old}}+h$\;
                $m_{\text{new}} \gets \tanh{(I)}$,\quad $m_{\text{new},i-1} \gets j$\;
                $\epsilon \gets \left(\sum_{i}|m_{\text{new,i}}-m_{\text{old,i}}|\right)/\left(\sum_{i}|m_{\text{new,i}}+m_{\text{old,i}}|\right)$\; 
                $m_{\text{old}}\gets \lambda m_{\text{new}} + (1-\lambda)m_{\text{old}}$\;
        }
        
           $m_{\text{avg}} \gets m_{\text{new}}$  \Comment*[r]{spin averages}
           $p(m_k=\pm 1) \gets 1/\left(1+\exp{(\mp2I_k)}\right)$ \Comment*[r]{individual probabilities}
        $p(m_k=\pm1|m_i=j)\gets 1/\left(1+\exp{(\mp2I_k)}\right)$ \Comment*[r]{conditional probabilities}
    }
}
$\langle m_i m_j \rangle \leftarrow$ Compute correlations from Eq.~(\ref{eq:correlation_basic})
%\tcc{correlations}
%\For{$i \gets 1$ \KwTo $N $}{
    % \For{$j \gets 1$ \KwTo $N$}{
%         $\langle m_im_j\rangle \gets p(m_i=-1|m_j=-1)p(m_j=-1)-p(m_i=-1|m_j=1)p(m_j=1)-p(m_i=1|m_j=-1)p(m_j=-1)+p(m_i=1|m_j=1)p(m_j=1)$
%     }
% }
\end{algorithm}
\vspace{-25pt}

\section{Experiments}
\begin{figure}[!h]
    \vspace{-5pt}
    \centering
    \includegraphics[width=1\textwidth,keepaspectratio]{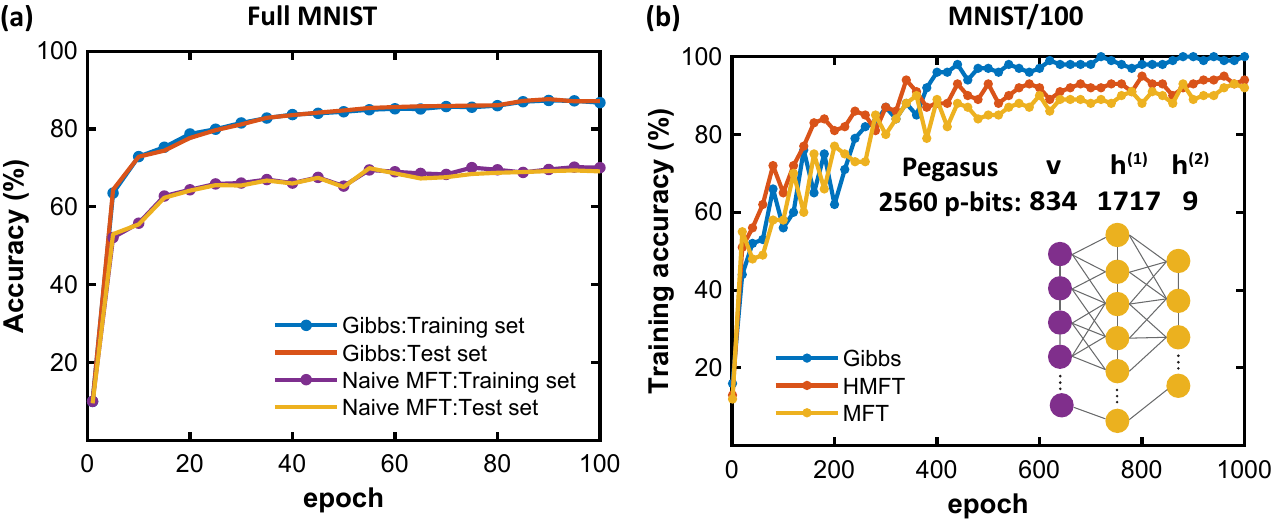}
    %\vspace{-15pt}
    \caption{\textbf{MNIST accuracy with different methods:} (a) Full MNIST (60,000 images) is trained on sparse DBM (Pegasus 2560 p-bits) with Gibbs sampling (CD-$10^{5}$) and naive MFT where batch size = 50, learning rate = 0.003, momentum = 0.6. Around 87\% accuracy is achieved in 100 epochs for Gibbs sampling and 70\% for the naive MFT. Test accuracy represents the accuracy of all 10,000 images from the MNIST test set, while the training accuracy corresponds to the accuracy of 10,000 images randomly drawn from the training set. (b) Training accuracy of MNIST/100  with the three different schemes: naive MFT, HMFT, and Gibbs sampling where they perform similarly. Here the batch size = 10, momentum = 0.6 and learning rate varies from 0.06 to 0.006 over 1000 epochs.}
    \label{fig:accuracy}
\end{figure}

\vspace{-8pt}
We have used the MNIST dataset (handwritten digits, \cite{lecun1998mnist}) to train sparse, deep and unrestricted Boltzmann networks without any downsampling, typically performed on hardware implementations by D-Wave and others \cite{adachi2015application,manukian2019accelerating,dixit2021training,bohm2022noise}. We have used black/white images by thresholding the MNIST dataset and we choose a Pegasus graph  \cite{dattani2019pegasus} with up to 2560 p-bits (nodes) as the sparse DBM network model in this paper. The graph density of this Pegasus is 0.55\% and the maximum number of neighbors is 15. The network has 834 visible p-bits (including 5 sets of labels each containing 10 p-bits) and 1726 hidden p-bits that are arranged in 2 layers as shown in the inset of FIG.~\ref{fig:accuracy}b. The total number of connections (network parameters) in this graph is 17984. Using our fast Gibbs sampler (p-computer), we accomplish the contrastive divergence algorithm to train the MNIST dataset divided into 1200 mini-batches with 50 images in each batch. We used $10^5$ sweeps in the negative phases of each epoch.

Similarly, we train the full MNIST using our hybrid MFT algorithm with naive MFT in the positive phase and Gibbs sampling ($10^5$ sweeps) in the negative phase. To find the classification accuracy, we perform a softmax classification over 50 label p-bits to get the 10 labels. The p-bit with the highest probability of being `1' indicates the classified digit. FIG.~\ref{fig:accuracy}a shows that our sparse DBM with 2560 p-bits reaches around 87\% accuracy with Gibbs sampling and 70\% accuracy (with MFT tolerance $=10^{-2}$ and this accuracy may further improve with lower tolerance) with hybrid MFT technique in 100 epochs despite having a significantly lesser number of parameters than typical RBMs. For the full MNIST dataset, the computational expense of HMFT prevented us from comparing it with the results of MFT at this time but in future this could be made possible with more parallel resources like using GPUs. Moreover, sophisticated techniques like layer-by-layer learning \cite{salakhutdinov2009deep} should further improve the accuracy reported in this work. Although our reported accuracy is comparable with models with less parameters (e.g., regression models), the real value of Boltzmann machines is their generative properties and has been shown recently in \cite{niazi2023training}.

%Similarly, we train the full MNIST using our hybrid MFT algorithm with naive MFT in the positive phase and Gibbs sampling ($10^5$ sweeps) in the negative phase. To find the classification accuracy, we perform a softmax classification over 50 label p-bits to get the 10 labels. The p-bit with the highest probability of being `1' indicates the classified digit. FIG.~\ref{fig:accuracy}a shows that our sparse DBM with 2560 p-bits reaches around 87\% accuracy with Gibbs sampling and 70\% accuracy with hybrid MFT technique in 100 epochs despite having a significantly lesser number of parameters than typical RBMs. For the full MNIST dataset, the computational expense of HMFT prevented us from comparing it with the results of MFT.

To be able to evaluate both the efficiency of MFT and HFMT methods in the positive phase, we performed the simpler task of training  MNIST/100 (100 images randomly chosen from the MNIST dataset). Three different schemes used the same hyper-parameters where the positive phase of training is accomplished with naive MFT, HMFT (on CPU), and Gibbs sampling (on p-computer). The negative phase is performed in our probabilistic computer where we are naturally doing persistent CD algorithm (PCD) \cite{hinton2012practical,tieleman2008training}. This hybrid computing scheme is illustrated in Fig.~\ref{fig:fig2} and the details of the experimental setup can be found in \cite{niazi2023training}. The training accuracy of 100 images reaches 100\% with Gibbs sampling and naive MFT and HMFT also perform similarly. Supplementary Table~\ref{tab:benchmarking2} indicates that despite a large difference in the training set log-likelihood, the test set shows roughly similar results, indicating how the hybrid approach does not degrade the performance significantly. The performance of this approach on larger datasets and networks remains to be seen.

It is interesting to note here that when Gibbs sampling in the negative phase is replaced by xMFTs, both training and test set accuracies degrade severely and, in fact, do not work at all. The supplementary Table~\ref{tab:benchmarking1} shows the poorer performance of xMFTs in the negative phases. 
\vspace{-15pt}
\section{Conclusions and Outlook}
\vspace{-6pt}
The end of Moore's law is driving the development of physics-based probabilistic computers that can accelerate MCMC algorithms by orders of magnitude. In this paper, we showed an FPGA emulation of such a computer that can take up to 45 billion Gibbs samples per second. Experimentally-validated projections indicate up to $10^{15}$ samples per second are possible with truly physical p-computers. Such a large increase in MCMC speeds may allow the direct training of deep and \emph{unrestricted} BMs. As an example of this approach, we trained a 2-layer unrestricted deep BM on a sparse Pegasus graph used by D-Wave, showing promising results (near 90\% classification accuracy with only $\approx$18k parameters). To aid with the positive phase training of unrestricted and deep BMs, we also proposed a hierarchical mean-field theory assisted learning algorithm. In accordance with common wisdom, we found that MFTs fail to estimate model correlations, especially when the weights become large. Surprisingly, however, we observed that MFTs accurately approximate data correlations, during the positive phase, greatly simplifying training in hardware. With the development of new physical computers, our results may allow the training of deep and unrestricted BMs previously thought to be intractable where hybrid MFT approaches could be used during pretraining or as a supplement to the computationally expensive Gibbs sampling.  Extensions of the model by fully exploiting the parallelism offered by the MFTs in deeper networks to train harder datasets are left for future study.

\vspace{-8pt}
\acksection
\vspace{-8pt}
Authors acknowledge support from the Office of Naval Research Young Investigator Program and National Science Foundation grants. 
\clearpage 

% \setcounter{algocf}{0}%

% % please don't delete the following lines
% \SetKwComment{Comment}{\normalfont $\triangleright$ }{}
% \SetKwInOut{Input}{Input}
% \SetKwInOut{Output}{Output}
% \SetKwInput{KwSampler}{Sampler}
% \SetKw{KwBy}{By}
% \begin{algorithm}[!ht]
% \DontPrintSemicolon
% \caption{\textbf{The Hierarchical Mean-field Algorithm}}
% \label{alg:alg3}
% \Input{weights and biases, update factor, tolerance, maximum iteration}
% %\KwSampler{Classic Gibbs CPU, Graph-colored Gibbs CPU, Graph-colored Gibbs FPGA}
% \Output{estimates for averages and correlations }
% \For{each p-bit}{
%     solve basic mean-field equations to estimate average\;
% }
% \For{each p-bit}{
%     clamp it to $+1$\;
%     solve basic mean-field equations for all other p-bits\;
%     compute probability of $\pm1$ for all other p-bits\;
%     \;
%     clamp it to $-1$\;
%     solve basic mean-field equations for all other p-bits\;
%     compute probability of $\pm1$ for all other p-bits\;
% }
% compute correlations from hierarchical MFT\;
% \end{algorithm}

%\bibliographystyle{unsrtnat}
%\bibliography{references}

\section*{Supplementary Information}
\beginsupplement

\setcounter{section}{0}

\section{HMFT vs NMFT in a toy 2-spin example}
\label{sec:toy_example}
Consider a simple two-spin antiferromagnetic system with weights  $W_{ij}=\delta_{ij}-1$ and no biases. At any temperature, the marginals of these two spins will be identically, i.e., $\langle m_i \rangle=0$. As such, the naive mean field method which uses conditional independence assumption $\langle m_i m_j\rangle = \langle m_i \rangle \langle m_j \rangle $ will estimate the correlation between these two spins to be identically zero, which is clearly incorrect, since at any non-zero temperature the spins will be anti-correlated. On the other hand, the hierarchical mean field method which \emph{does not} assume the conditional independence and rather uses Eq.~(4) to compute correlations, estimates a non-zero correlation.
For example, at $T=1$, the exact correlation (from Boltzmann law) between the two spins for the given weights and biases is $-0.7616$. The hierarchical method also estimates the same correlation value. Higher-order correlations (hierarchically obtained) can also be shown to be better estimated by the HMFT method. 
\vspace{-2pt}
\section{Contrastive divergence algorithm}
In this section, we briefly outline the contrastive divergence algorithm in our hybrid approach which implements Eq.~(\ref{eq:CD}):
% please don't delete the following lines
\SetKwComment{Comment}{\normalfont $\triangleright$ }{}
\SetKwInOut{Input}{Input}
\SetKwInOut{Output}{Output}
\SetKwInput{KwSampler}{Sampler}
\SetKw{KwBy}{By}
\begin{algorithm}
\caption{\textbf{Mean-field assisted training of sparse DBMs}}
\label{alg:alg2}
\Input{number of samples $N$,  batch size $B$, number of batches $N_{B}$, epochs $N_{\text{L}}$, learning rate $\varepsilon$, mean-field update factor $\lambda$, mean-field tolerance $\delta$, maximum iteration for mean-field $T_{\mathrm{max}}$}
%\KwSampler{Probabilistic Computer (FPGA}
\Output{trained weights $J_{\text{out}}$ and biases $h_{\text{out}}$}
$J_{\text{out}} \gets \mathcal{N}(0,0.01)$,\quad $h_{\text{out,hidden}} \gets 0$,\quad $h_{\text{out,visible}} \gets \log{(p_i/(1-p_i))}$\;
\For{$i\gets 1$ \KwTo $N_{\text{L}}$}{
    \For{$j \gets 1 $ \KwTo $N_{\text{B}}$}{
        %$J_{\text{Sampler}} \gets J_{\text{out}}$\;
        \tcc{positive phase}
        \For{$k \gets 1$ \KwTo $B$}{
            $h_{\text{$B$}} \gets \text{ clamping to batch images}$\;
            %$h_{\text{Sampler}} \gets h_{\text{$B$}}+h_{\text{out}}$\;
            %$\{m\} \gets \text{Sampler}(N)$ \Comment*[r]{p-computer}\ 
            $\langle m_i\rangle^{(k)}, \langle m_im_j\rangle^{(k)} \gets 
            \text{\textbf{x}MFT$\_$module}(J_{\text{out}},h_{B},\lambda,\delta,T_{max})$\;
            }
            $\langle m_i\rangle_{\text{data}} = \text{mean}\left(\{\langle m_i\rangle^{(k)}\}\right)$, \quad $\langle m_im_j\rangle_{\text{data}} = \text{mean}\left(\{\langle m_im_j\rangle^{(k)}\}\right)$ \Comment*[r]{CPU}
        \tcc{negative phase}
        $h_{\text{Sampler}} \gets h_{\text{out}}$, \quad $J_{\text{Sampler}} \gets J_{\text{out}}$\;
        $\{m\} \gets \text{GibbsSampler}(N)$ \Comment*[r]{p-computer}
        $\langle m_i\rangle_{\text{model}} = \text{mean}(\{m\})$, \quad $\langle m_im_j\rangle_{\text{model}} = \{m\}\{m\}^{\text{T}}/(N)$\Comment*[r]{CPU}
        \tcc{update weights and biases} 
        $J_{\text{out},ij} \gets J_{\text{out},ij} + \epsilon\left(\langle m_{i}m_j\rangle_{\text{data}}-\langle m_{i}m_j\rangle_{\text{model}}\right)$\ \Comment*[r]{CPU}\ 
        $h_{\text{out},i} \gets h_{\text{out},i} + \epsilon\left(\langle m_{i}\rangle_{\text{data}}-\langle m_{i}\rangle_{\text{model}}\right)$ \Comment*[r]{CPU}
    }
}
\end{algorithm}
\vspace{-10pt}
\section{Evolution of correlations over epochs}
\begin{figure}[!h]
    \vspace{0pt}
    \centering
    \includegraphics[width=0.75\textwidth,keepaspectratio]{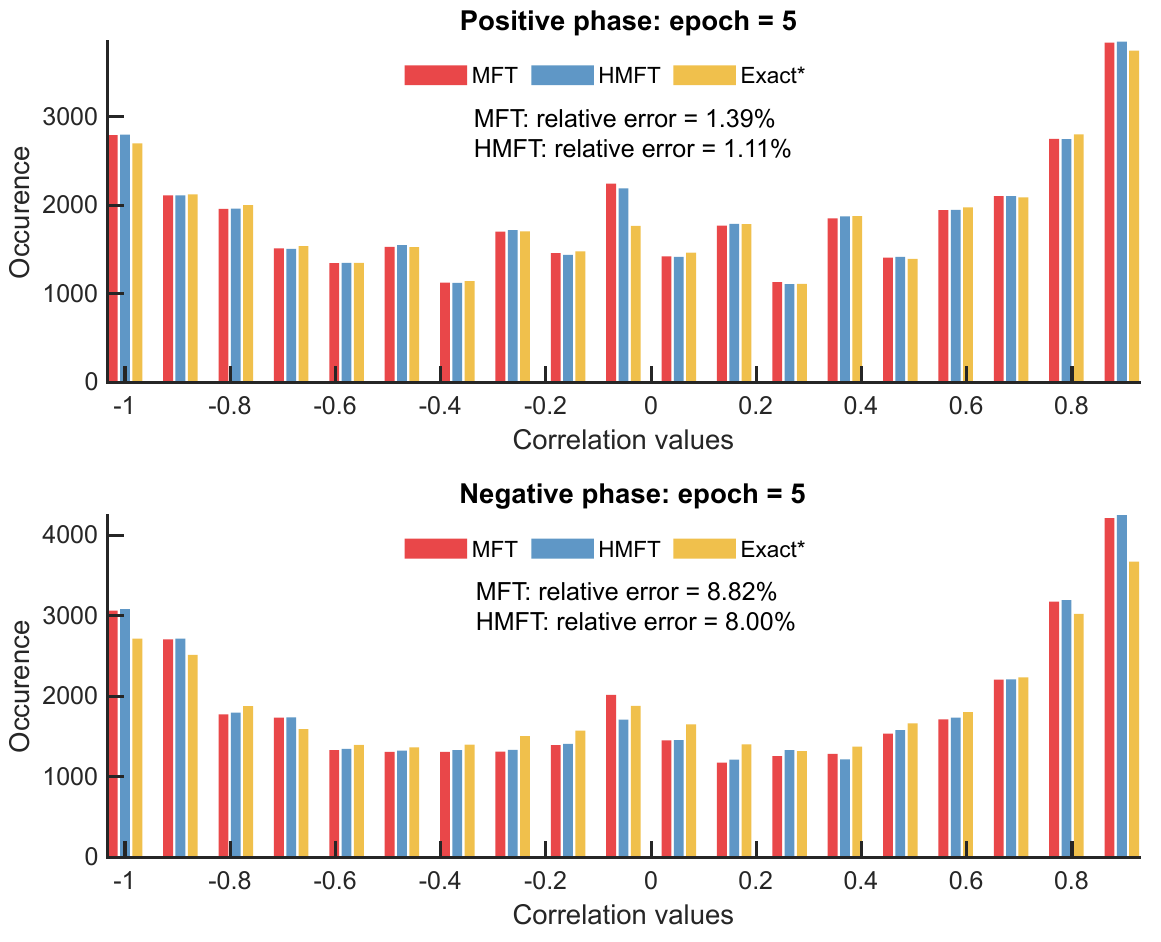}
    \includegraphics[width=0.8\textwidth,keepaspectratio]{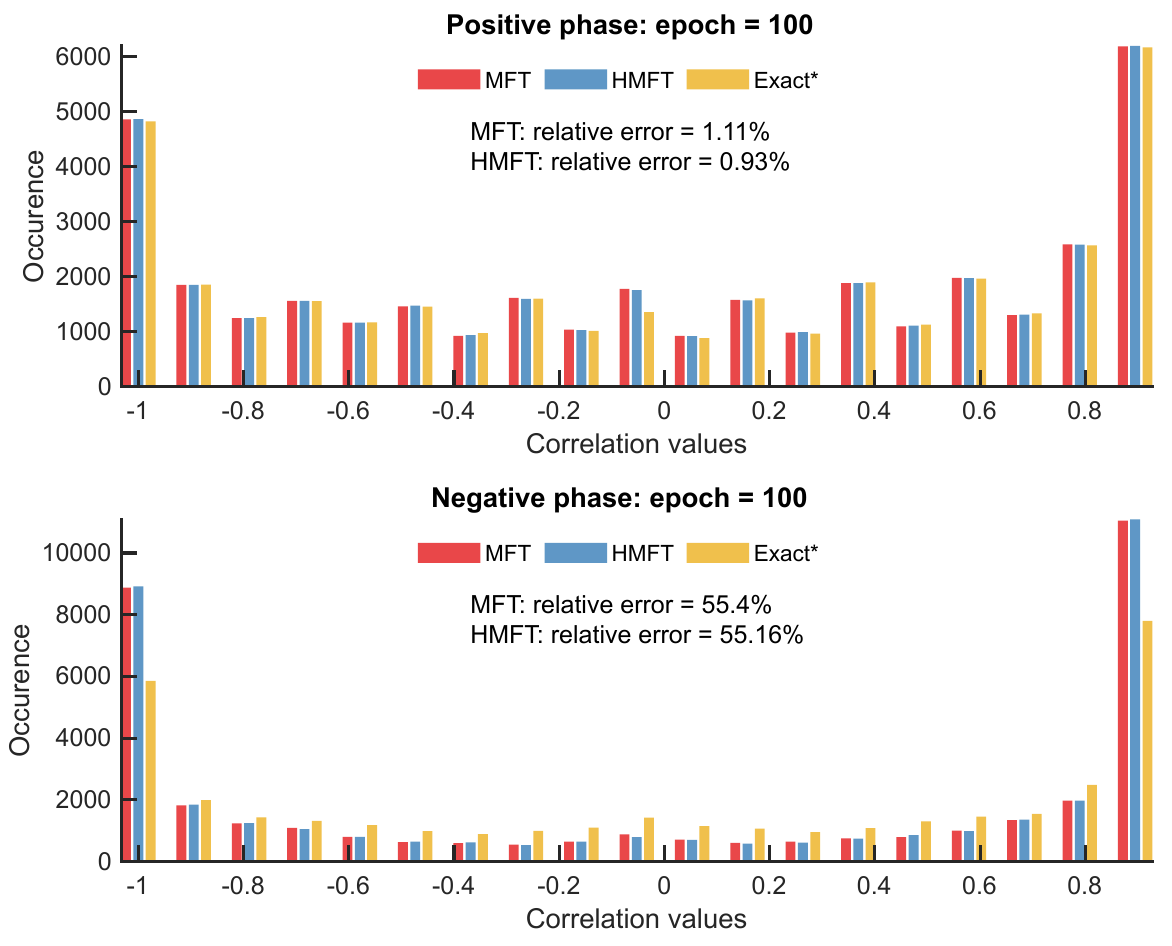}
    \caption{\textbf{Typical predictions of correlations from different methods:} The correlations predicted by the three different schemes - naive MFT, HMFT, and Gibbs sampling (the putative exact method since we cannot obtain exact Boltzmann correlations in general spin-glasses)  are shown during a typical epoch in the training of a sparse DBM. We used a batch size of 10 images and for the positive phase, we obtained correlations by showing only one batch of 10 images. For Gibbs sampling, we used $10^4$ sweeps in both positive and negative phases. We chose a relative error tolerance of $10^{-2}$ for both MFT and HMFT. 20 bins were used in both histograms. \textbf{MFT algorithms do significantly better in the positive phase than in the negative phase} allowing their use in the positive phase training of deep and unrestricted BMs, instead of the more expensive Gibbs sampling.}
    \label{fig:fig4}
\end{figure}

In this section, we discuss the evolution of correlations over epochs for naive and hierarchical MFT. Typical predictions of correlations from xMFT algorithms are shown in Supplementary Fig.~\ref{fig:fig4}. It can be clearly seen that in the positive phase, xMFT algorithms provide nearly accurate estimations for correlations. The clamping of many pbits during the positive phase helps xMFT algorithms to boost their performance but in the absence of such clamps, their performances degrade severely in the negative phase. In a hybrid setting, the probabilistic computer suffers from the communication delay caused by the fact that images are to be sent repeatedly whereas in the negative phase, there is no such delay. These two considerations justify the desire to replace Gibbs sampling in the positive phase with xMFT algorithms.

In order to provide a quantitative measure of the performance between the two xMFT algorithms discussed in this paper, we define the relative average error between the two correlation matrices estimated from two different approaches as $
\epsilon=\sqrt{\sum_{ij}{(A_{ij}-B_{ij})^2}}/\sqrt{\sum_{ij}{B_{ij}^2}}$
where $A$ is the correlation matrix predicted by the xMFT algorithms and $B$ is the corresponding matrix for the Gibbs sampling. We do not include the zero correlation values in this measure. Since at the 2560 p-bits level, it is impossible to obtain correlations from the exact distribution, we use Gibbs sampling as the ``putative" reference for comparison. Supplementary Table~\ref{tab:benchmarking1} lists this measure for the xMFT algorithms both in the positive and negative phases. The performance of the xMFT algorithms in both phases is consistent with Supplementary Fig.~\ref{fig:fig4}. As mentioned in the main text, in our HMFT implementations we do not modify the average estimations from naive MFT therefore both xMFT algorithms show the same accuracy for averages. For correlations, HMFT estimates are slightly better than naive MFT. It is also interesting to note that both models (MFT/HFMT) perform better at lower epochs when the weights are small which may suggest their use in possible pre-training supplementing the exact Gibbs sampling approach. 

\vspace{-5pt}
\begin{table*}[h]
    \centering
     \caption{{\footnotesize Relative average error is shown in positive and negative phases for two MFT algorithms. The error is measured with respect to Gibbs sampling at each phase (we take $10^4$ sweeps for both positive and negative phases). The tolerance of the MFT algorithms is set to $10^{-2}$. xMFT predicted averages and correlations are significantly better in the positive phase than in the negative phase.}}
     \vspace{4pt}
    \begin{tabular}{@{}lcccccccc@{}}
        \toprule
   
          &\multicolumn{4}{c}{\bf Positive phase} & \multicolumn{4}{c}{\bf Negative phase}  \\
        \midrule
           \textbf{epoch} & \multicolumn{2}{c}{\bf averages } & \multicolumn{2}{c}{\bf correlations} & \multicolumn{2}{c}{\bf averages} & \multicolumn{2}{c}{\bf correlations} \\
           & \multicolumn{2}{c}{$\langle m_i \rangle$ } & \multicolumn{2}{c}{ $\langle m_im_j \rangle$ } &  \multicolumn{2}{c}{$\langle m_i \rangle$ } & \multicolumn{2}{c}{ $\langle m_im_j \rangle$ } \\
         \midrule
         & {\textbf{MFT}} & {\textbf{HMFT}} &  {\textbf{MFT}} & {\textbf{HMFT}} & {\textbf{MFT}} & {\textbf{HMFT}} & {\textbf{MFT}} & {\textbf{HMFT}}\\
         \midrule
          1 & 0.72\% & 0.72\% & 1.55\% & 1.19\% & 2.00\% & 2.00\% & 3.82\% & 2.967\% \\
          5 & 0.69\% & 0.69\% & 1.39\% & 1.11\% & 5.82\% & 5.82\% & 8.82\% & 8.00\%\\
          10 & 0.63\% & 0.63\% & 1.24\% & 1.00\% & 11.73\% & 11.73\% & 16.40\% & 16.07\%\\
          50 &  0.59\% & 0.59\% & 1.11\% & 0.92\% & 35.85\% & 35.85\% & 46.80\% & 46.61\%\\
          100 & 0.61\% & 0.61\% & 1.11\% & 0.93\% & 44.44\% & 44.44\% & 55.4\% & 55.16\% \\  
        \bottomrule
    \end{tabular}
    \label{tab:benchmarking1}
\end{table*}

\vspace{-12pt}
\section{Log-likelihood measure for xMFT algorithms}
\vspace{-12pt}
\begin{table*}[h]
    \centering
     \caption{{\footnotesize Train (100 images) and test set (20 images) log-likelihood for different samplers i.e., Gibbs-Gibbs,  naive MFT-Gibbs, HMFT-Gibbs for sparse deep Boltzmann machines.}}
     \vspace{4pt}
    \begin{tabular}{@{}lcccc@{}}
        \toprule
        {\bf MNIST/100} & {\bf Gibbs-Gibbs}  & {\bf HMFT-Gibbs} & {\bf MFT-Gibbs} \\
        \midrule
        Train set  & -35.08  &  -71.07 & -89.82 \\
        Test set  & -33.87  &  -37.71  & -37.63\\
        
        \bottomrule
         \end{tabular}
   \label{tab:benchmarking2}
\end{table*}
In this section, we report the log-likelihood ($\mathcal{L}$) measure defined as
\begin{equation}
\mathcal{L} = \sum_{i=1}^{N} \sum_{c=1}^{C} y_{i, c} \cdot \log(p_{i, c})
\end{equation}
for the xMFT algorithms in both the training and the test set. Here, $y_{i, c}$ is the true label of the $i^{\text{th}}$ sample for class $c$ (1 if the sample belongs to class $c$ and 0 otherwise), and $p_{i, c}$ is the predicted probability that the $i^{\text{th}}$ sample belongs to class $c$. We measure the performance after training MNIST/100, with three different choices of algorithms to be used in the positive phase namely, Gibbs sampling, naive MFT and HMFT. The negative phase is always performed with Gibbs sampling. Supplementary Table~\ref{tab:benchmarking2} lists this measure.

As can be seen, the HMFT-trained model performs better than the naive MFT-trained model in both the training and test set although in the test set case, this difference is minimal.

\end{document}